\documentclass[twocolumn,superscriptaddress,aps,preprintnumbers,amsmath,amssymb,prl,nofootinbib]{revtex4-1}

\usepackage{graphicx}
\usepackage{epstopdf}
\usepackage{dcolumn}
\usepackage{bm}
\usepackage{hyperref}
\usepackage{color}

\begin{document}


\def\a{\alpha}
\def\b{\beta}
\def\c{\varepsilon}
\def\d{\delta}
\def\e{\epsilon}
\def\f{\phi}
\def\g{\gamma}
\def\h{\theta}
\def\k{\kappa}
\def\l{\lambda}
\def\m{\mu}
\def\n{\nu}
\def\p{\psi}
\def\q{\partial}
\def\r{\rho}
\def\s{\sigma}
\def\t{\tau}
\def\u{\upsilon}
\def\v{\varphi}
\def\w{\omega}
\def\x{\xi}
\def\y{\eta}
\def\z{\zeta}
\def\D{\Delta}
\def\G{\Gamma}
\def\H{\Theta}
\def\L{\Lambda}
\def\F{\Phi}
\def\P{\Psi}
\def\S{\Sigma}

\def\o{\over}
\def\beq{\begin{eqnarray}}
\def\eeq{\end{eqnarray}}
\newcommand{\gsim}{ \mathop{}_{\textstyle \sim}^{\textstyle >} }
\newcommand{\lsim}{ \mathop{}_{\textstyle \sim}^{\textstyle <} }
\newcommand{\vev}[1]{ \left\langle {#1} \right\rangle }
\newcommand{\bra}[1]{ \langle {#1} | }
\newcommand{\ket}[1]{ | {#1} \rangle }
\newcommand{\EV}{ {\rm eV} }
\newcommand{\KEV}{ {\rm keV} }
\newcommand{\MEV}{ {\rm MeV} }
\newcommand{\GEV}{ {\rm GeV} }
\newcommand{\TEV}{ {\rm TeV} }
\newcommand{\1}{\mbox{1}\hspace{-0.25em}\mbox{l}}
\newcommand{\headline}[1]{\noindent{\bf #1}}
\def\diag{\mathop{\rm diag}\nolimits}
\def\Spin{\mathop{\rm Spin}}
\def\SO{\mathop{\rm SO}}
\def\O{\mathop{\rm O}}
\def\SU{\mathop{\rm SU}}
\def\U{\mathop{\rm U}}
\def\Sp{\mathop{\rm Sp}}
\def\SL{\mathop{\rm SL}}
\def\tr{\mathop{\rm tr}}
\def\mpl{M_{PL}}

\def\IJMP{Int.~J.~Mod.~Phys. }
\def\MPL{Mod.~Phys.~Lett. }
\def\NP{Nucl.~Phys. }
\def\PL{Phys.~Lett. }
\def\PR{Phys.~Rev. }
\def\PRL{Phys.~Rev.~Lett. }
\def\PTP{Prog.~Theor.~Phys. }
\def\ZP{Z.~Phys. }

\def\dd{\mathrm{d}}
\def\ff{\mathrm{f}}
\def\BH{{\rm BH}}
\def\inf{{\rm inf}}
\def\ev{{\rm evap}}
\def\eq{{\rm eq}}
\def\SM{{\rm sm}}
\def\Mpl{M_{\rm Pl}}
\def\GeV{{\rm GeV}}
\newcommand{\Red}[1]{\textcolor{red}{#1}}


\title{
Simple realization of inflaton potential on a Riemann surface}

\author{Keisuke Harigaya}
\affiliation{Kavli IPMU (WPI), TODIAS, University of Tokyo, Kashiwa, 277-8583, Japan}
\author{Masahiro Ibe}
\affiliation{ICRR, University of Tokyo, Kashiwa, 277-8582, Japan}
\affiliation{Kavli IPMU (WPI), TODIAS, University of Tokyo, Kashiwa, 277-8583, Japan}
\begin{abstract}
The observation of the B-mode in the cosmic microwave background radiation combined with the
so-called Lyth bound suggests the trans-Planckian variation of the inflaton field during inflation.
Such a large variation generates concerns over inflation models 
in terms of the effective field theory below the Planck scale.
If the inflaton resides in a Riemann surface and the inflaton potential is a multivalued function of the inflaton field
when it is viewed as a function on a complex plane, the Lyth bound can be satisfied 
while keeping field values in the effective field theory within the Planck scale.
We show that a multivalued inflaton potential can be realized starting from a single-valued Lagrangian 
of the effective field theory below the Planck scale.
\end{abstract}

\date{\today}
\maketitle
\preprint{IPMU 14-0095}
\preprint{ICRR-report-678-2014-4}

\headline{Introduction}\\
Cosmic inflation scenario~\cite{Guth:1980zm} has succeeded in not only solving the flatness and the horizon problems
in the standard cosmology, but also providing the origin of the large scale structure 
of the universe and the fluctuation of the cosmic microwave background (CMB) radiation~\cite{Mukhanov:1981xt}.
Precise observations of the CMB have revealed the nature of inflation~\cite{Hinshaw:2012aka,Ade:2013uln}, 
and the observations can be well-explained by slow-roll inflation~\cite{Linde:1981mu,Albrecht:1982wi}.

Recently, the BICEP2 collaboration reported a large tensor fraction in the CMB, 
$r=O(0.1)$~\cite{Ade:2014xna}.
Such a large tensor fraction is known to require a variation of the inflaton field value larger than the Planck scale 
during inflation (the so-called Lyth bound~\cite{Lyth:1996im}).
Thus, the observed tensor fraction apparently indicates a trans-Planckian inflaton field value during inflation.
The chaotic inflation model~\cite{Linde:1983gd}, which shows a perfect fit with
the BICEP2 results, is an excellent example in which the inflaton field value is much larger
than the Planck scale.

For obvious reasons, however, models with such a large field value
seem to be highly 
sensitive to the physics beyond the Planck scale 
including the theory of quantum gravity such as string theory.
Without having any knowledge on the quantum gravity, the field theory is at the best 
considered to be an effective theory whose Lagrangian is given by 
a series expansion in fields.
In particular, higher dimensional operators suppressed by the Planck scale encode
the effects of the quantum gravity over which we have no control
without knowing the nature of the fundamental theory.

One of the
way out to tame
the higher dimensional operators
is to assume an approximate shift symmetry
in the fundamental theory,
so that the potential is almost unchanged 
by the shift of the inflaton field~\cite{Freese:1990rb,Kawasaki:2000yn,Kallosh:2010ug,Harigaya:2014qza}.
The shift symmetry also guarantees the flatness of the inflaton potential.

In this letter, we discuss an alternative way out where fields  appearing in a series expansion of the 
effective field theory never exceed the Planck scale while the  inflaton field satisfies the Lyth bound.
There, the inflaton resides in a Riemann surface and the inflaton potential is a multivalued function 
of the inflaton field when it is viewed as a function on a complex plane.
In this case, we can realize a field variation much larger than the Planck scale while keeping its amplitude
within the Planck scale during inflation.
This viewpoint has been considered in the context of the axion monodromy~\cite{Silverstein:2008sg,McAllister:2008hb,Shlaer:2012by} motivated in string theory and its field theoretical approaches in Refs.~\cite{Kaloper:2008fb,Berg:2009tg,Kaloper:2011jz,Kaloper:2014zba}.
As we will see, we show that such a multivalued inflaton potential and an inflaton field residing in a Riemann surface 
can be realized starting from a single-valued Lagrangian,
where the appearance of the effectively enhanced field space is obtained by charge assignments 
of fields which break a $U(1)$ symmetry.%
\footnote{See also alternative possibilities to 
realize the effective trans-Planckian inflaton in terms of fields within the Planck scale 
by aligning several potentials of natural inflation~\cite{Kim:2004rp} 
or by using the collective behavior of multi-inflatons~\cite{Dimopoulos:2005ac,Ashoorioon:2009wa}. 
}

\vspace{0.5cm}

\headline{Lyth Bound}\\
Here, let us briefly review the so-called Lyth bound~\cite{Lyth:1996im}.
The magnitude of the tensor perturbation depends only on the inflation scale during inflation.
On the other hand, the magnitude of the scalar perturbation not only depends 
on the inflation scale but also on the slow-roll-ness of the inflaton field during inflation.
As the inflaton rolls faster, the effect of the quantum fluctuation of the inflaton on the scalar perturbation is suppressed, since the scalar perturbation is essentially the fluctuation of the inflationary period.
Therefore, for a given inflation scale, the tensor fraction becomes larger for a 
faster rolling of the inflaton.

As a result,  there is a lower-bound on the variation of the inflaton field during inflation 
$\Delta \phi$, the so-called Lyth bound~\cite{Lyth:1996im};
\begin{eqnarray}
\label{eq:Lyth bound}
\frac{\Delta \phi}{M_{ PL}} \gtrsim 1.6 \times\left(\frac{r}{0.2}\right)^{1/2},
\end{eqnarray}
where $M_{PL}\simeq 2.4 \times 10^{18}$ GeV denotes the reduced Planck scale.%
\footnote{Here, we have taken the $e$-folding number to be $\D N_e \simeq 10$.}
Therefore, the observed tensor fraction, $r\simeq 0.2$, indicates the inflaton field value much 
larger than the Planck scale.
As discussed above, such a large field value generates concerns over inflation models 
based on the effective field theory where the Lagrangian is given by a series expansion in fields.

\vspace{0.5cm}
\headline{Multivalued Inflaton Potential}\\
A possible way around the argument of the Lyth bound is to assume that
the inflaton potential is a multivalued function of the inflaton when
the inflaton is considered to reside in a complex plane.
As a simple example, let us imagine that the inflaton potential is given by 
\begin{eqnarray}
V \propto \phi^{1/N} + h.c.\ ,
\end{eqnarray}
where $N$ is an arbitrary integer.
Obviously, this potential is multivalued when it is viewed as a function on a complex plane.
Then, let us further assume that the inflaton field $\phi$ is a complex scalar field 
which takes a value not on a simple complex plane but on a Riemann surface for $\phi^{1/N}$, i.e.
\begin{eqnarray}
\phi = |\phi| \times e^{i\alpha}~~(\alpha = 0-2N\pi)\ .
\end{eqnarray}
With this assumption, the inflaton potential is now a single-valued function
on the  inflaton Riemann surface, and hence, we can achieve a field variation much larger than 
the field amplitude itself, i.e.
\begin{eqnarray}
\Delta \phi = \left|\textstyle{\int} d\phi \right| \gg |\phi|\ ,
\end{eqnarray}
for $N\gg 1$.%
\footnote{
The usage of the multivalued inflaton potential is crucial since otherwise the 
inflaton potential shows a trivial periodicity during the large variation $\D\phi$,
which has no physical meaning.
}

\begin{figure}[t]
 \begin{center}
  \includegraphics[width=0.98\linewidth]{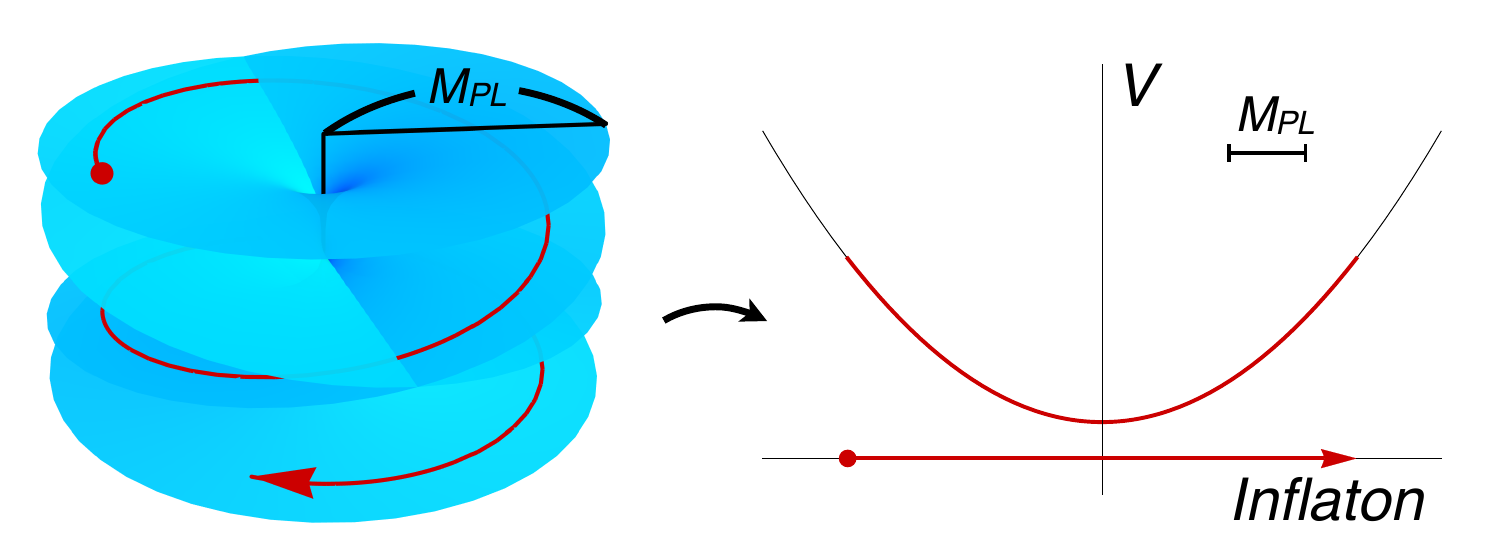}
 \end{center}
\caption{\sl A schematic picture of the inflaton potential on a Riemann surface.
Left) A Riemann surface for the function $\phi^{1/4}$.
We assume that the inflaton field takes a value on this surface.
Right)
Inflaton potential $(V \propto \phi^{1/4})$
on the Riemann surface.
}
\label{fig:potential}
\end{figure}

In Fig.\,\ref{fig:potential}, we show a schematic picture to illustrate the inflaton field residing in a Riemann surface 
for $N=4$.
As the figure shows, the variation of the inflaton field can be much larger than the amplitude.
On the Riemann surface, the multivalued inflaton potential becomes single-valued.
In this way, we can construct a large field inflation model effectively.

One of serious drawbacks of this idea is, however, that the multivalued potential has a singularity 
at the origin of the field space.
Such a singularity is far from acceptable from the viewpoint of the 
effective field theory below the Planck scale, where we assume that 
the effective Lagrangian is given by a series expansion in terms of regular fields.
Therefore, the question is whether we can construct a model with a multivalued 
inflaton potential starting from a single-valued field theory.

\vspace{0.5cm}
\headline{Effective Multivalued Inflaton Potential}\\
To construct a model with a multivalued inflaton potential
out of a single-valued potential, let us consider two complex scalar fields $\phi$ and $S$ which reside in 
complex planes.
We also assume a continuous global $U(1)$ symmetry%
\footnote{
We may also replace the continuous symmetry with a
discrete symmetry such as $Z_M$ with a large integer $M$.
}
with the charge assignments 
\begin{eqnarray}
\phi: N,~~S : 1,
\end{eqnarray}
where $N$ is a sufficiently large integer.
The scalar potential consistent with the $U(1)$ symmetry is given by
\begin{eqnarray}
\label{eq:scalar potential}
V &=& V(\phi \phi^*, S S^*, \phi^* S^N)\nonumber\\
&=& -m_{\phi}^2 \phi \phi^* + y_\phi (\phi \phi^*)^2 - m_{S}^2 S S^* + y_S (S S^*)^2\nonumber\\
&& +\left( \frac{c}{M_{PL}^{N-3}}\phi^* S^N + h.c \right) + \cdots,
\end{eqnarray}
where $m_{\phi}^2,$ $m_S^2$, $y_\phi$, $y_S$ and $c$ are constants and $\cdots$ denotes higher dimensional terms irrelevant for our discussion.

We assume that $\phi$ and $S$ obtain non-vanishing vacuum expectation values (VEVs), which is realized by negative masses-squared of $\phi$ and $S$ around the origin.
Then the radial directions of $\phi$ and $S$, and one linear combination of the phase directions of them obtains large masses from the scalar potential in Eq.~(\ref{eq:scalar potential}).
However, another linear combination of the phase directions of $\phi$ and $S$, which corresponds to a Nambu-Goldstone boson (NGB) associated with spontaneous breaking of the global $U(1)$ symmetry, remains massless.

Along the NGB direction, the field values of $\phi$ and $S$ are related by
\begin{eqnarray}
\label{eq:vev relation}
S = \lambda \phi ^{1/N},
\end{eqnarray}
due to the $U(1)$ symmetry.
Here, $\lambda$ is a constant determined by the scalar potential.
It should be emphasized here that the phase of $S$ is ``multivalued" function
of the phase of $\phi$, which plays a crucial role in the following discussion.
We note that since the charge of $\phi$ is larger than that of $S$, the NGB is mostly the phase direction of $\phi$ if the VEVs of $\phi$ and $S$ are of the same order, which we assume in the following.

Now let us explicitly break the $U(1)$ symmetry softly by introducing a potential,
\begin{eqnarray}
\label{eq:UVpotential}
\Delta V = \Lambda^3 S + {\rm h.c.},
\end{eqnarray}
where $\Lambda$ is an order parameter of the explicit breaking of the $U(1)$ symmetry.
We assume that the scale $\Lambda$ is sufficiently smaller than the VEVs of $\phi$ and $S$.
Then the low energy effective theory is well-described by the (now pseudo-)NGB with 
the explicit breaking term in Eq.~(\ref{eq:UVpotential}).
Since the pseudo-NGB is mostly composed of the phase direction of $\phi$,
the dynamics of the NGB is approximately identified with the dynamics of the phase of $\phi$.
Expressing the potential by $\phi$ using Eq.~(\ref{eq:vev relation}), the resulting low energy effective potential along the pseudo-NGB is given by
\begin{eqnarray}
\label{eq:effective}
\Delta V_{\rm eff} =\lambda \Lambda^3 \phi^{1/N} + {\rm h.c.}\ ,
\end{eqnarray}
which is nothing but the multivalued inflaton potential discussed in the previous section.
The phase direction (i.e. the pseudo-NGB) plays the role of the inflaton.

\begin{figure}[tb]
\begin{center}
  \includegraphics[width=.7\linewidth]{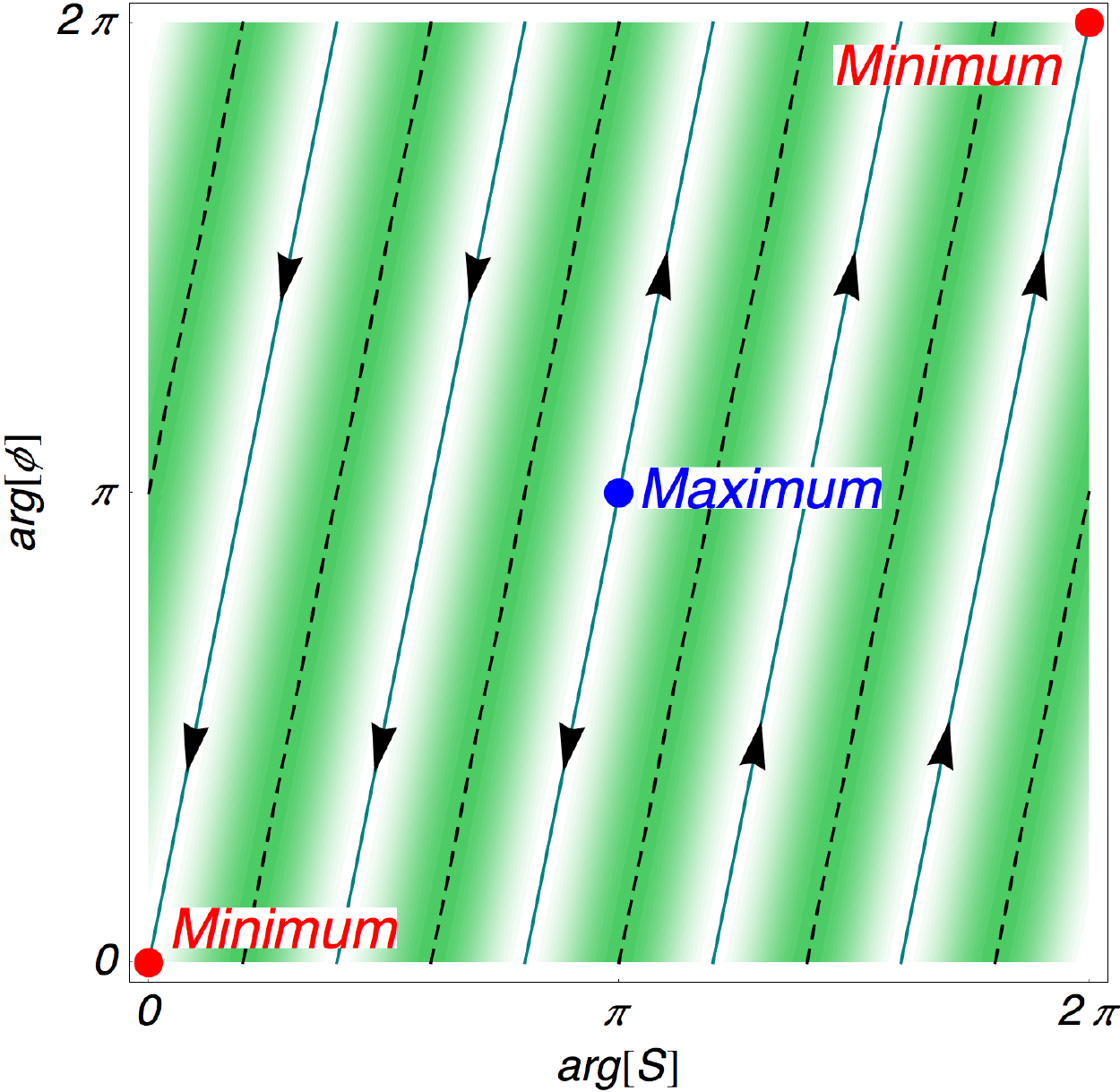}
 \end{center}
\caption{\sl \small
An inflaton trajectory on phases of $S$ and $\phi$ for $N=5$.
The inflaton potential in Eq.\,(\ref{eq:effective}) is realized along 
the trajectory shown by the solid line.
Arrowheads indicate the height of the potential along the trajectory,
and the field values which maximize and minimize the potential in Eq.\,(\ref{eq:effective}) 
are denoted by points.
Here, we have shifted the phases of $\phi$ and $S$ so that ${\rm arg} \phi = {\rm arg} S=0$ is the minimum of the potential. 
The shaded regions have large potentials by the last term in Eq.~(\ref{eq:scalar potential}).
}
\label{fig:lock}
\end{figure}

The inflaton dynamics with the multivalued potential can be understood in the following way.
Due to the relation in Eq.~(\ref{eq:vev relation}), when $\phi$ rotates $2\pi$, $S$ rotates only $2\pi/N$.
Then, since the inflaton potential is provided for the phase of $S$ as in Eq.\,(\ref{eq:UVpotential}),
the periodicity of the potential for the phase of $\phi$, which is the main component of the inflaton, is effectively enlarged to $2\pi N$.
This non-trivial periodicity is the origin of the effective multivalued nature and the trans-Planckian variation of the inflaton field during inflation.
In Fig.\,\ref{fig:lock}, we show the potential on the phases of $S$ and $\phi$ for $N=5$.
The inflaton trajectory corresponds to the bottom of the valley along which the potential is very flat 
over the range of the $[0,2\pi N]$.

To show the inflaton potential quantitatively, let us extract a canonically normalized pseudo-NGB, $a$, given by an identification,
\begin{eqnarray}
\label{eq:NGB}
\phi &\rightarrow& \vev{\phi} {\rm exp}\left[i N \frac{a}{f_a}\right],~~
S \rightarrow \vev{S} {\rm exp}\left[ i  \frac{a}{f_a}\right],\nonumber\\
f_a &\equiv& \sqrt{2 N^2 |\vev{\phi}|^2 + 2 |\vev{S}|^2}.
\end{eqnarray}
The scalar potential of $a$ is given by
\begin{eqnarray}
\label{eq:potential1}
V(a) = 2 |\Lambda^3 \vev{S}| \left( 1- {\cos}\frac{a}{f_a}\right),
\end{eqnarray}
where we have eliminated a constant phase and a sign inside the cosine by shifting $a$.
Here, we have added a constant term to the potential so that the cosmological constant vanishes at the vacuum.

As a result, we obtain the potential of the so-called natural inflation~\cite{Freese:1990rb}, 
which is consistent with the results of the Planck~\cite{Ade:2013uln} and the BICEP2~\cite{Ade:2014xna} 
experiments for $f_a \gsim 5 \mpl$~\cite{Freese:2014nla}.%
\footnote{For a consistency of more generalized natural inflation models with the BICEP2 and the Planck results, see e.g.~\cite{Czerny:2014qqa}.}
If $N$ is sufficiently large, $f_a$ can be as large as $5 M_{PL}$ while keeping $\phi$ and $S$ within the Planck scale,
which is possible due to the multivalued nature of the effective potential in Eq.\,(\ref{eq:effective}).

Before closing this section, let us discuss how small VEVs are acceptable in this model.
For that purpose, let us remember that the two scalar fields are connected 
via higher dimensional operators in Eq.~(\ref{eq:scalar potential}),
\begin{eqnarray}
\label{eq:connect}
V \supset \frac{c}{M_{PL}^{N-3}}\phi^* S^N + h.c. \ ,
\end{eqnarray}
without which we have two $U(1)$ symmetries.
When this operator is ineffective, the linear combination of the phase directions other then the NGB, $b$,  becomes lighter than $a$.
In this case, $b$ plays a role of the inflaton, whose decay constant in the potential is given by $f_a/N$.
Therefore, the variation of the inflaton field $b$ cannot exceed the Planck scale.
As a result, it is required that
\begin{eqnarray}
\vev{\phi} \simeq \vev{S} \gg M_{PL} \left( \frac{2 N^2}{c}
\left(\frac{m_{\rm inf}}{M_{PL}}\right)^2
\right)^{1/(N-1)}\ .
\end{eqnarray}
Here, we have expressed the condition in terms of the mass of the inflaton $m_{\rm inf}$.
It is determined by the normalization of the CMB fluctuation;
$m_{\rm inf} = {\cal O}(10^{13})$\,GeV for $f_a \gg M_{PL}$.
Thus, we find that the VEVs of $S$ and $\phi$ are bounded from below,
\begin{eqnarray}
\vev{\phi} \simeq \vev{S} \gg 10^{-10/(N-1)}M_{PL} \ .
\end{eqnarray}

\vspace{0.5 cm}
\headline{Multivalued Axion Inflation}\\
Instead of putting the explicit breaking of the $U(1)$ symmetry, the potential in Eq.~(\ref{eq:UVpotential}), 
we may generate the breaking by non-perturbative dynamics.
Let us assume QCD-like gauge dynamics which exhibits spontaneous breaking of chiral symmetries at a scale 
$\Lambda_{\rm dym}$ far below the VEVs of $\phi$ and $S$.
We couple the field $S$ to the gauge dynamics via a Yukawa coupling;
\begin{eqnarray}
{\cal L}_{\rm int} = y S Q \bar{Q}, 
\end{eqnarray}
where $Q$ and $\bar{Q}$ are fermion fields charged under the gauge symmetry,
and $y$ denotes a coupling constant.

Now that the $U(1)$ symmetry has an anomaly of the gauge symmetry,
as is the case with the QCD-axion\,\cite{Peccei:1977hh,Weinberg:1977ma,Wilczek:1977pj}, 
the pseudo-NGB obtains a potential,
\begin{eqnarray}
V &\simeq& m_f \Lambda_{\rm dyn}^3 \left( 1- {\rm cos}\left({\rm arg}S  \right)\right)\nonumber\\
&=& m_f \Lambda_{\rm dyn}^3 \left( 1- {\rm cos}\left({\rm arg}\phi^{1/N} \right)\right),
\end{eqnarray}
where we have eliminated constant phases.
Here, we have assumed that there is a fermion charged under the gauge symmetry with a mass $m_f < \Lambda_{\rm dyn}$.
We have again obtained a multivalued potential of the field $\phi$.

In terms of the canonically normalized NGB $a$, 
the scalar potential of $a$ is given by
\begin{eqnarray}
V(a) =  \Lambda_{\rm dyn}^4 \left( 1- {\rm cos}\frac{a}{f_a}\right),
\end{eqnarray}
which again leads to the potential of the natural inflation~\cite{Freese:1990rb}.
As emphasized above, the multivalued nature played a crucial role to realize $f_a \gg M_{PL}$
while keeping the VEVs of $S$ and $\phi$ below the Planck scale.

\vspace{0.5cm}
\headline{Discussion}\\
We have considered inflation models such that the inflaton potential is a multivalued 
function when it is viewed as a function on a complex plane
while the inflaton field (effectively) resides in a Riemann surface.
In this way, we can satisfy the Lyth bound while fields appearing in the effective field theory
are sub-Planckian.
We have shown some simple examples where the multivalued inflaton potential
is realized starting from a single-valued Lagrangian of the effective field theory below the Planck scale.

It should be emphasised that the effectively enhanced inflaton field space originates from
the charge assignments of fields which break a $U(1)$ symmetry.
This mechanism should be contrasted with other attempts to realize the effectively 
trans-Planckian field variation by, for example, alignment between several potentials of natural 
inflation~\cite{Kim:2004rp} or by using the collective behavior of multi-inflatons~\cite{Dimopoulos:2005ac,Ashoorioon:2009wa}.  
In our model, on the other hand, we can realize the trans-Planckian field variation
using only two fields without having alignments.

As we have shown, the simple examples result in the natural inflation model.
There, the shift symmetry, which is often imposed to control the inflaton potential~\cite{Freese:1990rb,Kawasaki:2000yn,Kallosh:2010ug,Harigaya:2014qza}, is understood as non-linear realization of a global $U(1)$ symmetry.

So far, we have imposed the global $U(1)$ symmetry.
Since the symmetry is explicitly broken to generate 
the multivalued inflaton potential, the origin of the symmetry should be scrutinized in future works.
It would be interesting to construct a model such that the $U(1)$ symmetry is an accidental one
as a result of some other well-motivated symmetries.

\vspace{1 cm}

%
\headline{Acknowledgements}\\
This work is supported by
Grant-in-Aid for Scientific research 
from the Ministry of Education, Science, Sports, and Culture (MEXT), Japan, No. 24740151 and 25105011 (M.I.),
from the Japan Society for the Promotion of Science (JSPS), No. 26287039 (M.I.),
the World Premier International Research Center Initiative (WPI Initiative), MEXT, Japan (K.H. and M.I.)
and a JSPS Research Fellowships for Young Scientists (K.H.).
%

\end{document}